\documentclass[a4paper,11pt]{article}

\usepackage{amsmath}
\usepackage{graphicx}
\usepackage{pstricks}
\newcommand{\lt}{\left}
\newcommand{\rt}{\right}

\newcommand{\alp}{\alpha}

\newcommand{\omg}{\omega}

\newcommand{\lmd}{\lambda}
\newcommand{\tha}{\theta}

\newcommand{\vph}{\varphi}

\newcommand{\Omg}{\Omega}

\newcommand{\Lmd}{\Lambda}

\newcommand{\half}{\frac{1}{2}}

\newcommand{\nab}{\nabla}

\newcommand{\beqn}{\begin{eqnarray}}
\newcommand{\eeqn}{\end{eqnarray}}
\newcommand{\be}{\begin{equation}}
\newcommand{\ee}{\end{equation}}
\newcommand{\benn}{\begin{equation*}}
\newcommand{\eenn}{\end{equation*}}
\newcommand{\bi}{\begin{itemize}}
\newcommand{\ei}{\end{itemize}}
\newcommand{\bpm}{\begin{pmatrix}}
\newcommand{\epm}{\end{pmatrix}}
\author{ Hungsoo Kim$^*$, H. W. Lee$^*$, Y. S.
Myung \footnote{ Relativity Research Center, Inje University,
Gimhae 621-749, Korea } \footnote{ Institute of Theoretical
Science, 5203 University of Oregon, Eugene, OR 97403, USA}
\footnote{Email-address : ysmyung@inje.ac.kr}}

\title{Role of the  Brans-Dicke scalar in the holographic description of dark energy}
\begin{document} 
\maketitle
\begin{abstract}
We study cosmological application of the holographic energy
density in the Brans-Dicke theory. Considering the holographic
energy density as a dynamical cosmological constant, it is more
natural to study it in the Brans-Dicke theory than in  general
relativity. Solving the Friedmann and Brans-Dicke field equations
numerically, we clarify the role of Brans-Dicke field during
evolution of the universe. When the Hubble horizon is taken as the
IR cutoff, the equation of state  ($w_{\Lmd}$) for the holographic
energy density is  determined to be $\frac{5}{3}$  when the
Brans-Dicke parameter $\omg$ goes infinity. This means that the
Brans-Dicke field  plays a crucial role in  determining the
equation of state. For the particle horizon IR cutoff, the
Brans-Dicke scalar mediates a transition from $w_{\Lmd} = -1/3$
(past) to $w_{\Lmd} = 1/3$ (future). If a dust matter is present,
it determines future equation of state. In the case of future
event horizon cutoff, the role of the Brans-Dicke scalar and dust
matter are turned out to be  trivial, whereas the holographic
energy density plays an important role as a dark energy candidate
with $w_{\Lmd} =-1$.
\end{abstract}

\section{Introduction}
Type Ia supernova obervations\cite{SuperNova} suggest that our
universe is in accelerating phase and the dark energy contributes
$\Omg_{DE} \simeq 0.60-0.70$ to the critical energy density of the
present universe. Also cosmic microwave background
observations\cite{CMB} imply that the standard cosmology is given
by inflation and FRW universe \cite{StdCosmology}.

A typical candidate for the dark energy is the cosmological
constant in general relativity. Recently Cohen {\em et
al}\cite{holographic} showed that in the effective theory of
quantum field theory, the UV cutoff $\Lmd$ is related to the IR
cutoff $L_{\Lmd}$, due to the limit set by forming a black hole.
In other words, if $\rho_{\Lmd}$ is the quantum zero-point energy
density caused by the UV cutoff, the total energy density of the
system with size $L_{\Lmd}$ should not exceed the mass of the
system-sized black hole : $L_{\Lmd}^3 \rho_{\Lmd} \leq
L_{\Lmd}/G$. Here the Newtonian constant $G$ is related to the
Planck mass by  $G = 1/M_p^2$. The largest IR cutoff  $L_{\Lmd}$
is chosen as the one saturating this inequality and the
holographic energy density is then given by $\rho_{\Lmd} = 3c^2
M_p^2/8\pi L_{\Lmd}^2$ with an appropriate  factor $3c^2/8\pi$.
Comparing with the cosmological constant, we regard it as a
dynamical cosmological constant. Taking $L_{\Lmd}$ as the size of
the present universe (Hubble horizon $R_{HH}$), the resulting
energy density is comparable to the present dark energy density
\cite{Horava}. Even though this holographic approach leads to the
data, this description is incomplete because it fails to explain
the equation of state for the dark energy-dominated
universe\cite{Hsu}. In order to resolve this situation, one
introduces other candidates for the IR cutoff. One is the particle
horizon $R_{PH}$. This provides $\rho_{\Lmd} \sim a^{-2(1+1/c)}$,
which means that the equation of state is given by $\omg_{\Lmd} =
1/3$ for $c = 1$\cite{Miao}. However, it corresponds to a
radiation-dominated universe and  it is a decelerating phase. In
order to find an accelerating phase, we need to introduce the
future event horizon $R_{FH}$. In the case of $L_{\Lmd} = R_{FH}$,
one finds $\rho_{\Lmd} \sim a^{-2(1 - 1/c)}$ which could describe
the dark energy with $\omg_{\Lmd} = -1$ for $c = 1$. This is close
to the data\cite{SuperNova} and the related works appeared in
ref.\cite{HH,Myung,Medved,ON}.

On the other hand, it is worthwhile to investigate the holographic
energy density in the framework of the Brans-Dicke theory. The
reasons are as follows. Because the holographic energy density
belongs to a dynamical cosmological constant, we need a dynamical
frame to accommodate it instead of general relativity. Further,
taking $L_{\Lmd}=R_{HH}$, it  fails to determine  the equation of
state $w_{\Lmd}$ in the general relativity framework.  In addition
to these, the Brans-Dicke scalar speeds up the expansion rate of a
dust matter-dominated era (reduces deceleration), while slows down
the expansion rate of cosmological constant era (reduces
acceleration)\cite{LS,Hongsu}. The Brans-Dicke generalization was
first studied by Gong\cite{Gong}. Since the Brans-Dicke
description of gravitation is to replace the Newtonian constant
$G$ by a time varying scalar $\Phi(t)$, the holographic energy
density is given by $\rho_{\Lmd} = 3 \Phi/8\pi L^2_{\Lmd}$ with
$c^2=1$. Gong recovered the same results as those in general
relativity for a large $\omg$.  The present authors studied  the
same issue by considering a Bianchi identity as a consistency
condition\cite{KLM}. The equation of state for Hubble IR cutoff
 is determined to be $w_{\Lmd}=\frac{5}{3}$  when the
Brans-Dicke parameter $\omg$ goes infinity. This implies that the
Brans-Dicke framework is suitable for studying an evolution of the
holographic energy density.

In this work, we introduce  a dust matter  to our consideration
and solve the equations numerically. Since the holographic energy
density is dynamical, it is nontrivial to solve the Friedmann  and
the Brans-Dicke field equation with three conservation laws.  They
can not be solved analytically. From this study we investigate the
role of the holographic energy density, Brans-Dicke scalar and
dust matter for a given IR cutoff. Especially, we wish to show why
a combination of the holographic energy density and future event
horizon could describe  a dark energy-dominated era.

\section{Brans-Dicke cosmology}
For cosmological purpose, we introduce the  Brans-Dicke (BD)
action  with a matter  \be
   S = \int d^4 x\sqrt{-g}
       \lt[
             \frac{1}{16\pi}
             \lt(
                   \Phi R - \omg \frac{\nab_{\alp}\Phi\nab^{\alp}\Phi}{\Phi}
             \rt)
            +{\cal L}_M
       \rt],
\ee where $\Phi$ is the BD scalar which plays the role of an
inverse of the Newtonian constant, $\omg$ is the parameter of BD
theory, and ${\cal L}_M$ represents other matter which takes a
perfect fluid form. The field equations for metric $g_{\mu\nu}$
and BD scalar $\Phi$ are \be\begin{split}
   G_{\mu\nu} &\equiv R_{\mu\nu} - \half g_{\mu\nu}R
               =      8\pi T_{\mu\nu}^{BD} + \frac{8\pi}{\Phi} T_{\mu\nu}^M, \\
   \nab_{\alp}\nab^{\alp}\Phi &= \frac{8\pi}{2\omg +
   3}{T^M}_{\alp}^{\;\alp},
\end{split}\ee
where the energy-momentum tensor for the BD scalar is defined by
\be
   T^{BD}_{\mu\nu}=
\frac{1}{8\pi}\Big[\frac{\omega}{\Phi^2}\Big(\nabla_{\mu}\Phi\nabla_{\nu}
\Phi-\frac{1}{2}g_{\mu\nu}(\nabla
\Phi)^2\Big)+\frac{1}{\Phi}\Big(\nabla_{\mu}\nabla_{\nu}\Phi-g_{\mu\nu}
\nabla_{\alpha}\nabla^{\alpha}\Phi\Big)\Big] \ee and the
energy-momentum tensor for other matter takes the form \be
   T^M_{\mu\nu} = p_M g_{\mu\nu} + (\rho_M + p_M)U_{\mu}U_{\nu}.
\ee Here $\rho_M(p_M)$ denote the energy density (pressure) of the
matter and $U_{\mu}$ is a four velocity vector with
$U_{\alp}U^{\alp} = 1$.

Assuming that our universe is homogeneous and isotropic, we work
with the Friedmann-Robertson-Walker (FRW) spacetime  \be
   ds^2 = -dt^2 + a^2(t)
                  \lt[
                        \frac{dr^2}{1-kr^2}
                      + r^2 \lt( d\tha^2 + \sin^2\tha d\phi^2 \rt)
                  \rt]. \quad
\ee We consider  a spatially flat spacetime of $k=0$. In the  FRW
spacetime, the field equations take the  forms \be\begin{split}
   H^2 + H\lt( \frac{\dot{\Phi}}{\Phi} \rt)
       - \frac{\omg}{6}\lt( \frac{\dot{\Phi}}{\Phi} \rt)^2
       = \frac{8\pi}{3}\frac{\rho_M}{\Phi},\\
   \ddot{\Phi} + 3H\dot{\Phi} = \frac{8\pi}{2\omg + 3}
                                 \lt( \rho_M - 3 p_M \rt)
\end{split}\ee
with the Hubble parameter $H = \dot{a}/a$. Here we note that the
case of $\omg=-3/2$ is not allowed when a matter with
$p_M\not=\rho_M/3$ comes into the BD theory.

Regarding the BD field as a perfect fluid, its energy and pressure
 are given by\cite{Hongsu}
  \be\begin{split}
   \rho_{BD} &= \frac{1}{16\pi G_0}
                \lt[
                      \omg \lt( \frac{\dot{\Phi}}{\Phi} \rt)^2
                    - 6H\frac{\dot{\Phi}}{\Phi}
                \rt], \\
   p_{BD}    &= \frac{1}{16\pi G_0}
                \lt[
                      \omg \lt( \frac{\dot{\Phi}}{\Phi} \rt)^2
                    + 4H\frac{\dot{\Phi}}{\Phi}
                    + 2\frac{\ddot{\Phi}}{\Phi}
                \rt],
\end{split}\ee
where $G_0$ is the present Newtonian  constant. Usually, if one
does not specify the parameter $\omg$, one cannot determine the BD
equation of state  exactly. The Bianchi identity leads to an
energy transfer between  BD field and  other matter \be
\label{BDC}
   \dot{\rho}_{BD} + 3H( \rho_{BD} + p_{BD} )
           = \frac{1}{G_0}\frac{\rho_M}{\Phi}\frac{\dot{\Phi}}{\Phi}
\ee and the  matter evolves  according to  its conservation law
\be \label{MC}\dot{\rho}_M + 3H( \rho_M + p_M ) = 0. \ee Their
equations of states are  given by \be
   w_{BD} \equiv \frac{p_{BD}}{\rho_{BD}}, \quad
   w_{M}  \equiv \frac{p_M}{\rho_M}.
\ee

\section{Brans-Dicke framework with holographic energy density and dust matter}
In this section, we investigate  how the equation of state for the
holographic energy  density changes when an interaction between
the BD field  $ \rho_{BD}$, holographic energy density $
\rho_{\Lmd}$ and dust matter $\rho_m$ is included. In this case
the Friedmann and BD field equations are \be\begin{split}
\label{FBD}
   H^2 + H\frac{\dot{\Phi}}{\Phi} - \frac{\omg}{6}
                                    \lt(
                                          \frac{\dot{\Phi}}{\Phi}
                                    \rt)^2
         = \frac{8\pi}{3}\frac{\rho_{t}}{\Phi},\\
   \ddot{\Phi} + 3H\frac{\dot{\Phi}}{\Phi}
         = \frac{8\pi}{2\omg+3}(\rho_{t} - 3p_{t}).
\end{split}\ee
Here $\rho_t = \rho_{\Lmd} + \rho_m$ and $p_t = p_{\Lmd} + p_m$.
The holographic energy density $\rho_{\Lmd}$ and a dust matter
$\rho_m$ are chosen to be  \be
   \rho_{\Lmd} = \frac{3}{8\pi}\frac{\Phi}{L_{\Lmd}^2}, \quad
   \rho_m = \rho_m^0 a^{-3}
\ee with $p_m = 0$.  In order to solve Eq.(\ref{FBD}) with
Eqs.(\ref{BDC}) and (\ref{MC}),  we define \be
   x = \ln a,\quad \vph = \frac{\Phi'}{\Phi},\quad \lmd = -\frac{H'}{H},\quad
   r = \frac{R'}{R},
\ee where $\prime$ means the derivative with respect to $x$ and $R
\in \{R_{HH},R_{PH},R_{FH}\}$. From the definition of $\vph$ and
$\lmd$, it is granted that $H$ and $\Phi$ are taken to be
positive\cite{Rendall}.  Then the Friedmann equation and BD field
equation become \be\begin{split}
   H^2\lt( 1 + \vph - \frac{\omg}{6}\vph^2 \rt) = \frac{8\pi}{3}
                                                  \frac{\rho_{t}}{\Phi},\\
   H^2\lt( \vph' - \lmd\vph + \vph^2 + 3\vph \rt) = \frac{8\pi}{2\omg + 3}
                                                    \frac{\rho_{t}-3p_{t}}
                                                         {\Phi}.
\end{split}\ee
 Also the
energy-momentum conservation law  leads to the pressure \be
   p_{\Lmd} = -\frac{1}{3}\lt( \vph - 2r + 3 \rt) \rho_{\Lmd} \ee whose  equation of state is given by \be
   w_{\Lmd} = -\frac{1}{3}(\vph - 2r +3).
\ee If other interaction between $\rho_{\Lmd}$ and $\rho_{m}$ is
included, then the equation of state $w_{\Lmd}$ takes a different
form\cite{myung2}.

\begin{figure}[h]
\centering
\includegraphics[width=10cm]{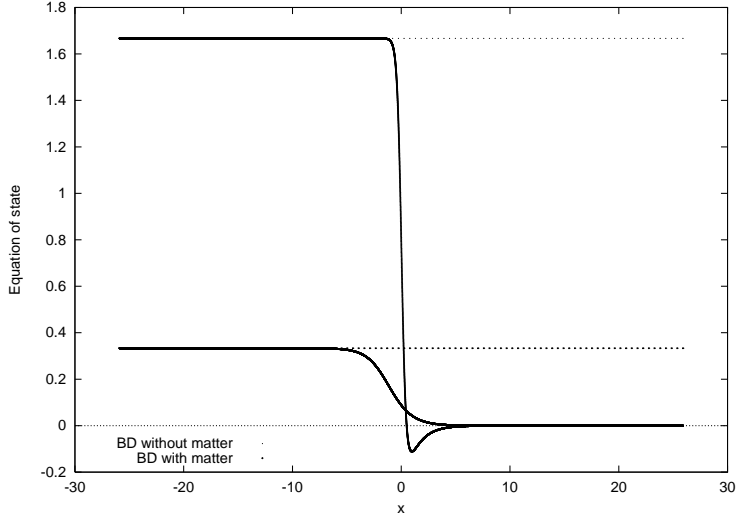}
\caption{A plot of  equation of state $w_{\Lmd}$ versus $x=\ln a$
for Hubble horizon. Here BD denotes the Brans-Dicke framework.}
\label{w_LMD_HH}
\end{figure}

From now on we focus on the change of $w_{\Lmd}$ by choosing
 an  IR cutoff $L_{\Lmd}$.
Firstly, we take Hubble horizon as the IR cutoff scale ($L_{\Lmd}
=R_{HH}= 1/H$). We have $\lmd = r$ and then eliminate $\lmd$ to
obtain \be\begin{split}
   \vph' &=  \frac{\omg (\omg + 1)\vph}{6}
             \Big(\vph - \frac{6}{\omg}\Big)
             \Big(\vph - \frac{1}{\omg +1}\Big),\\
       & r =  \half \Big(\frac{\omg \vph}{3} - 1\Big)
                \Big[
                      (\omg + 1)\vph -1
                \Big]
              + \frac{\vph + 3}{2}.
\end{split}\ee
One can solve the above equation numerically. We have a plot for
$w_{\Lmd}$ as is shown in Fig. \ref{w_LMD_HH}.
In Fig. \ref{w_LMD_HH}, an upper dotted line represents
 a graph for $w_{\Lmd}=\frac{5}{3}$ without a dust  matter in the limit of $\omg \to \infty$.
 This case was already found in\cite{KLM}.
Solid lines represent the change of  equation of state $w_{\Lmd}$
with a dust matter.  When a dust matter is present, the BD theory
allows two solutions in the far past:  $w_{\Lmd}=\frac{5}{3}$ for
a large $\omg$  and $w_{\Lmd}=\frac{1}{3}$, independent of $\omg$.
As the BD field evolves, the equation of state for the holographic
energy density converges that of dust matter. The universe behaves
as a dust matter-dominated phase  in the far future, irrespective
of where it starts. If the BD scalar is turned off, one cannot
determine the equation of state for the holographic energy density
only\cite{Hsu}. In this sense, although we do not obtain a dark
energy era,  the BD framework  is essential for determining
$w_{\Lmd}$ and it goes well with $L_{\Lmd} = 1/H$.

\begin{figure}[h]
\centering
\includegraphics[width=10cm]{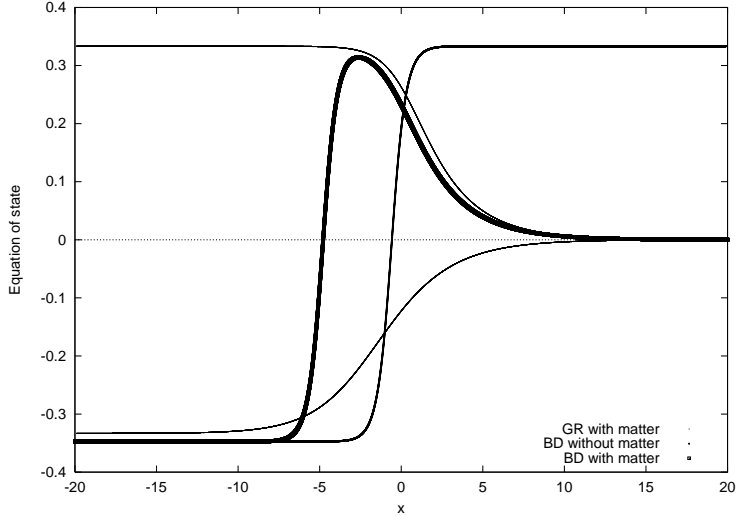}
\caption{A plot for $w_{\Lmd}$ as a function of $x=\ln a$ for
particle horizon. Here GR (BD) denote the general relativity
(Brans-Dicke) frameworks.} \label{w_LMD_PH}
\end{figure}

\begin{figure}[h]
\centering
\includegraphics[width=10cm]{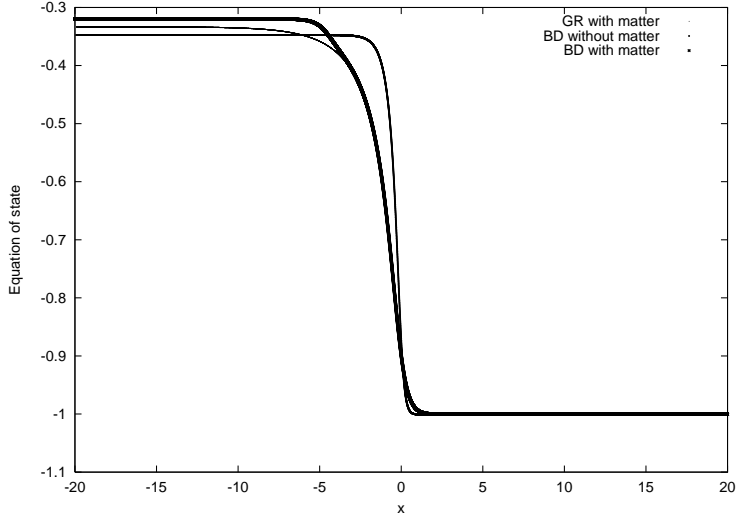}
\caption{A plot for $w_{\Lmd}$ as a function of $x=\ln a$ for
future event horizon. Here GR (BD) denote the general relativity
(Brans-Dicke) frameworks.} \label{w_LMD_FH}
\end{figure}

For particle horizon IR cutoff  with $L_{\Lmd}=R_{PH}\equiv
a\int^a_0\frac{da}{Ha^2}$ and future event horizon IR cutoff
$L_{\Lmd}=R_{FH}\equiv a\int^{\infty}_a \frac{da}{Ha^2}$, $r$ is
given by \be
   r = 1 \pm \sqrt{\Omg_{\Lmd}\Big(1 + \vph -
   \frac{\omg}{6}\vph^2\Big)}
\ee with $\Omg_{\Lmd} \equiv \rho_{\Lmd}/\rho_t$. Here $+$ denotes
particle horizon, while  $-$ represents future event horizon.
Eliminating $\lmd$ leads to  two coupled equations
\be\begin{split}
   \vph' &= -\frac{1 + \vph - \frac{\omg}{6}\vph^2}{2\omg + 3}
             \Big[
                   3\lt\{(\omg + 1)\vph - 1 \rt\}
                   +(\vph - 2r + 3)(\omg\varphi - 3)\Omg_{\Lmd}
             \Big],\\
   \Omg_{\Lmd}' &= (\vph - 2r + 3)\Omg_{\Lmd}(1 - \Omg_{\Lmd}).
\end{split}\ee
Here $\lmd$ is related to $r$ via \be
   \lmd =r+\frac{1 - \frac{\omg}{3}\vph}
                   {2(1 + \vph - \frac{\omg}{6}\vph^2)}\vph'.
\ee We solve the coupled equations numerically and plot $w_{\Lmd}$
in Fig. \ref{w_LMD_PH} and Fig. \ref{w_LMD_FH}  for particle
horizon and future event horizon, respectively.

For particle horizon IR cutoff $L_{\Lmd}=R_{PH}$,
$w_{\Lmd}=\frac{1}{3}$ is found for the holographic energy density
solely when using general relativity.  A thin line stands for
$w_{\Lmd}$ in the general relativity framework together with a
dust matter. A medium line results from the DB framework without
matter and a thick line represents $w_{\Lmd}$ in the  BD framework
with a dust matter. In the general relativity framework, the
equation of state of holographic energy density starts with
$w_{\Lmd}=\pm \frac{1}{3}$ in the far past and ends with
$w_{\Lmd}=0$ like as  a dust matter. However, in the BD framework,
equation of state of the holographic energy density starts with
$w_{\Lmd}=-\frac{1}{3}$ at the far past and then, transits to
$w_{\Lmd}=\frac{1}{3}$. This implies that without matter, the
holographic energy density becomes a radiation. In the BD
framework together with a dust matter, the equation of state of
holographic energy density starts with $w_{\Lmd}=-\frac{1}{3}$ in
the far past. The BD field makes a transition to a radiation phase
and finally, $w_{\Lmd}$ transits to a dust matter in the far
future. In this case, the dust matter determines a  future
equation of state for the holographic energy density. This means
that  a dust matter dominates in the holographic energy density
with particle horizon. Finally we mention that the BD scalar plays
a role of the mediator between $w_{\Lmd}=- \frac{1}{3}$ and
$w_{\Lmd}= \frac{1}{3}$.

For future event horizon cutoff $L_{\Lmd}=R_{FH}$, one finds
$w_{\Lmd}=-1$ with $c^2=1$  for $\rho_{\Lmd}$ only in the general
relativity framework. A thin line stands for a graph of $w_{\Lmd}$
with a dust matter in general relativity. A medium/thick lines
correspond  to the BD theory framework without/with a dust matter.
A general relativistic analysis was carried out by Li\cite{Miao}.
Equation of state of the holographic energy density starts with
$w_{\Lmd}=-\frac{1}{3}$ in the far past and becomes a cosmological
constant with $w_{\Lmd}=-1$ in the far future.  For this IR
cutoff, the holographic energy density serves as a dark energy and
leads to an accelerating era. As is shown in Fig. \ref{w_LMD_FH},
this feature persists  even in the BD framework with or without a
dust matter. This means that the holographic energy density goes
well with future event horizon $L_{\Lmd}=R_{FH}$. On the other
hand, the role of dust matter and BD scalar is trivial when
comparing with  the holographic energy density.

\section{Summary}
\begin{table}
 \caption{Summary for future equation of state. Here three combinations for  holographic energy density ($\rho_{\Lmd}$),
  Brans-Dicke scalar ($\rho_{BD}$),
  and dust matter ($\rho_{m}$) are evaluated
 for IR cutoff ($L_{\Lmd}$) as  Hubble horizon ($R_{HH}$), particle horizon($R_{PH}$)
 and future event horizon($R_{FH}$), respectively.}
\begin{tabular}{|c|c|c|c|}
   \hline
   matter  &  $R_{HH}$  & $R_{PH}$ & $R_{FH}$ \\ \hline

 $\rho_{\Lmd}+\rho_{BD}$   &  $w_{\Lmd}=5/3$ &  $w_{\Lmd}=1/3$ &  $w_{\Lmd}=-1$ \\
$\rho_{\Lmd}+\rho_{m}$    & $w_{\Lmd}=0$   &  $w_{\Lmd}=0$ &
$w_{\Lmd}=-1$\\
 $\rho_{\Lmd}+\rho_{BD}+\rho_{m}$ & $w_{\Lmd}=0$  &   $w_{\Lmd}=0$
  &  $w_{\Lmd}=-1$\\
  \hline
\end{tabular}
 \end{table}
  We study cosmological application of
the holographic energy density in the Brans-Dicke framework.
Considering the holographic energy density as a dynamical
cosmological constant, it is more natural to study it in the
Brans-Dicke theory than in general relativity. Solving the
Friedmann and Brans-Dicke field equations numerically, we
investigate the role of Brans-Dicke field during evolution of the
universe.

We summarize  future equation of state for the holographic energy
density in Table 1. When the Hubble horizon is taken as the IR
cutoff, the equation of state for the holographic energy density
($w_{\Lmd}$) is determined to be $\frac{5}{3}$  when the
Brans-Dicke parameter $\omg$ goes infinity. This means that the
Brans-Dicke scalar is crucial for determining the equation of
state when comparing to the case of $\rho_{\Lmd}$. However, if a
dust matter is turned on, its future equation of state is
determined  by $w_{\Lmd}=0$, irrespective of the presence of the
Brans-Dicke scalar. Actually, the equation of state  for
$\rho_{\Lmd}+\rho_{m}$ can be determined to be $w_{\Lmd}=0$ by the
Friedmann equation\cite{Hsu}.

For particle horizon IR cutoff, the Brans-Dicke scalar mediates
the transition from $w_{\Lmd} = -1/3$ (past) to $w_{\Lmd} = 1/3$
(future). However, if a dust matter is present, it determines
future equation of state. Hence a dust matter plays an important
role in the holographic description with particle horizon.

In the case of future event horizon cutoff, the role of the
Brans-Dicke scalar and dust matter are trivial, whereas the
holographic energy density plays an important role as a dark
energy candidate with $w_{\Lmd} =-1$.

Consequently, we find the major roles in the holographic
description of  an evolving universe: BD scalar in Hubble horizon;
dust matter in particle horizon; holographic energy density in
future event horizon. The BD scalar plays a role of the  mediator,
but it does not determine the future equation of state if a dust
matter or the holographic energy density is present.

{\bf Acknowledgments}

H. Kim and H. Lee were in part supported by KOSEF, Astrophysical
Research Center for the Structure and Evolution of the Cosmos. Y.
Myung was supported by the Korea Research Foundation Grant
(KRF-2005-013-C00018).

\end{document}